\documentclass[]{llncs}
\usepackage{amsmath}
\usepackage{amsfonts}
\usepackage[T1]{fontenc}
\usepackage{graphicx}
\usepackage{color}
\usepackage{url}
\usepackage[a]{esvect}
\usepackage{enumitem}
\usepackage[utf8]{inputenc}
\usepackage{cleveref}
\usepackage{algorithm}
\usepackage{algpseudocode}

\usepackage{wrapfig}

\let\llncssubparagraph\subparagraph
\let\subparagraph\paragraph
\usepackage[compact]{titlesec}
\let\subparagraph\llncssubparagraph

\algnewcommand{\LineComment}[1]{\State \(\triangleright\) #1}

\def\doubleblind{1}

\if\doubleblind1
\newcommand{\blinded}[1]{{\color{red} [blinded]}} 
\else 
\newcommand{\blinded}[1]{#1}
\fi

\begin{document}

\title{HOTVis: Higher-Order Time-Aware Visualisation of Dynamic Graphs}

\author{Vincenzo Perri\inst{1}\orcidID{0000-0002-4203-1177} \and
Ingo Scholtes \inst{2,1}\orcidID{0000-0003-2253-0216}}

\institute{Data Analytics Group, Department of Informatics (IfI), University of Zurich, Switzerland \and
Chair of Data Analytics, Faculty of Mathematics and Natural Sciences, University of Wuppertal, Germany}

\maketitle      

\begin{abstract}
Network visualisation techniques are important tools for the exploratory analysis of complex systems.
While these methods are regularly applied to visualise data on complex networks, we increasingly have access to time series data that can be modelled as temporal networks or dynamic graphs.
In dynamic graphs, the temporal ordering of time-stamped edges determines the \emph{causal topology} of a system, i.e., which nodes can, directly \emph{and indirectly}, influence each other via a so-called \emph{causal path}.
This causal topology is crucial to understand dynamical processes, assess the role of nodes, or detect clusters.
However, we lack graph drawing techniques that incorporate this information into static visualisations.
Addressing this gap, we present a novel dynamic graph visualisation algorithm that utilises higher-order graphical models of causal paths in time series data to compute time-aware static graph visualisations.
These visualisations combine the simplicity and interpretability of static graphs with a time-aware layout algorithm that highlights patterns in the causal topology that result from the temporal dynamics of edges.
\end{abstract}

\section{Introduction}

Network visualisation techniques are a cornerstone in the exploratory analysis of data on complex systems.
They help us recognize patterns in relational data on complex networks, such as, e.g., clusters or groups of well-connected nodes, hierarchical and core-periphery structures, or highly important nodes~\cite{Battista1994,Kaufmann2003,Noguchi2019}.
However, apart from knowing which elements are connected, we increasingly have information on when and in which chronological order connections occurred.
Sources of time-stamped data include social interactions, click stream data in the web, financial transactions, passenger itineraries in transportation networks, or gene regulatory interactions~\cite{holme2015}.
Despite these applications, visualising \emph{time-stamped network data} is still a challenge~\cite{Beck2017_Taxonomy,holme2015}.
Common approaches use sequences of snapshots, where each snapshot is a graph of the connections active in a time interval or at a point in time, to animate the evolution of the topology.
Such animations can help us gain a high-level understanding of temporal activities in dynamic graphs. 
However, they are complex and cognitively demanding, which makes it hard to recognise patterns that determine how nodes influence each other over time.
Moreover the application of graph drawing algorithms to temporal snapshot necessitates a coarse graining of time.
This introduces a major issue:
we lose information on the chronological ordering of links that determines so-called \emph{time-respecting} or \emph{causal paths}~\cite{Kempe2000_TemporalNetworks,holme2015}.
In a nutshell, for two time-stamped edges $(a,b; t_1)$ and $(b,c; t_2)$ that occur at times $t_1$ and $t_2$, a \emph{causal path} $\vv{abc}$ from node $a$ via node $b$ to node $c$ can only exist if edge $(a,b)$ occurs before $(b,c)$, i.e. if $t_1<t_2$.
If the ordering of edges is reversed, such a causal path does not exist, i.e. node $a$ cannot influence node $c$ via $b$, neither directly nor indirectly. 
This simple example highlights how the temporal ordering of edges gives rise to \emph{causal topologies}.
While two edges $(a,b)$ and $(b,c)$ in a static graph imply that a (transitive) path $\vv{abc}$ exists, the temporal ordering of edges in dynamic graphs can invalidate this assumption.
This has important implications for the modelling of epidemic processes, random walk and diffusion dynamics, centrality measures used to rank nodes, or clustering techniques. 
It calls for a new class of higher-order network modelling, analysis, and visualisation techniques~\cite{lambiotte2019}.
In a recent review on state-of-the-art temporal network analysis~\cite{holme2015}, Holme highlights a lack of visualisation techniques that (i) go beyond cognitively demanding animations, and (ii) consider the complex topology of causal paths in high-resolution time series data: \textit{ ``[\ldots] temporal networks lack the intuitive visual component of static networks. Probably this is a fundamental property that cannot be completely altered, but there should be better visualization methods than we have now. Highest on our wish list is a method that both simplifies some structures and keeps (at least some) of the time-respecting paths (maybe at the cost of not having time on the abscissa).''
} ~\cite{holme2015}, p. 23.

Addressing this gap, we develop a visualisation algorithm that incorporates information on causal paths in dynamic graphs into simple (static) visualisations.
Our contributions are: (i) we highlight a lack of time-aware graph visualisation techniques that respects the \emph{causal topology} resulting from the ordering of edges in high-resolution data on dynamic graphs; (ii) we develop a dynamic graph drawing algorithm that generalises force-directed layouts to high-dimensional De Bruijn graph models of causal paths~\cite{deBruijn1946,Scholtes2017}; (iii) we assess the quality of our visualisations in synthetic and empirical time series data, and show that they help to detect temporal clusters invisible in static visualisations, and identify important vertices with high temporal centrality; (iv) we provide an Open Source \texttt{python} implementation of our algorithm \cite{pathpy}.
Focusing on time-aware static visualisations that highlight temporal patterns neglected by existing techniques, we take a new approach to dynamic graph drawing.
Considering recent works on learning optimal higher-order graph models from rich time series data~\cite{Scholtes2017}, our work opens perspectives to combine machine learning and visual data mining.

\begin{figure*}[!ht]

\centering
     	\includegraphics[width=.8\textwidth]{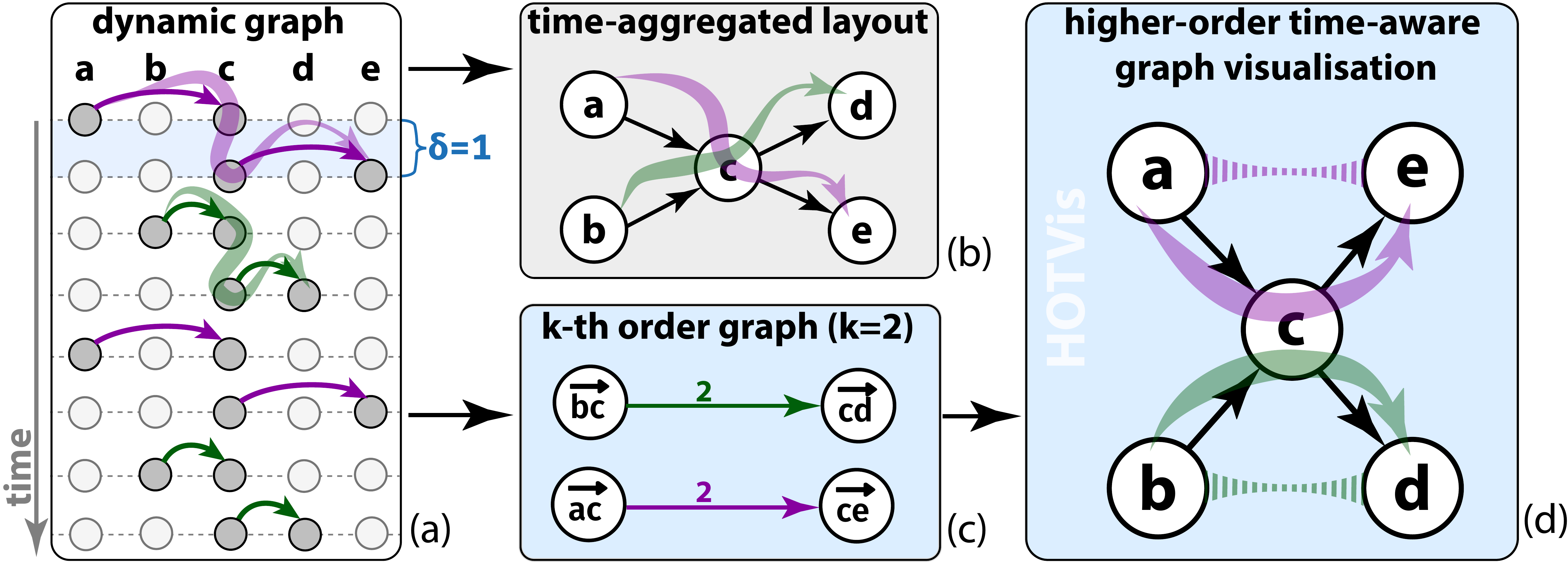}
        \caption{ 
        Information on \emph{causal paths} (coloured arrows) contained in dynamic graphs (a) is discarded by standard time-aggregated visualisations (b). \texttt{HOTVis} uses higher-order graph models of causal paths (c) to produce time-aware, static visualisations (d) that highlight the causal topology of dynamic graphs.}
        \label{fig:toyexample}
\end{figure*}

\section{Preliminaries and Related Work}
\label{sec:background}

\label{sec:background:temporalNets}

We define a dynamic graph as a tuple $G^{(t)}=(V,E^{(t)})$, where $V$ is a set of vertices and $E^{(t)}$ is a set of time-stamped edges $E^{(t)} \subseteq V \times V \times \mathbb{N}$.
We assume that $(v,w; t) \in E^{(t)}$ denotes that a directed edge between source vertex $v$ and target vertex $w$ occurred \emph{instantaneously} at discrete time $t \in \mathbb{N}$.
We say that a (static) graph $G=(V,E)$ is the \emph{time-aggregated} graph corresponding to a dynamic graph $G^{(t)}$ iff $(v,w) \in E \leftrightarrow \exists t \in \mathbb{N}: (v,w;t) \in E^{(t)}$.
We further assume that the edge weights $w:E \rightarrow \mathbb{N}$ of such a time-aggregated graph capture the number of times edges have been active in the corresponding dynamic graph, i.e. we define $w(v,w) := | \{ t \in \mathbb{N}: (v,w;t) \in E^{(t)} \}|$.
A simple example for a dynamic graph with eight time-stamped edges and five nodes is shown in \Cref{fig:toyexample} (a).
A key concept in the study of dynamic graphs is that of a \emph{time-respecting} or \emph{causal path}~\cite{Kempe2000_TemporalNetworks,holme2015}.
For a dynamic graph $G^{(t)}=(V,E^{(t)})$ we call a sequence $(v_0, v_1; t_0), (v_1, v_2; t_1)$, \ldots, $(v_{l-1}, v_l; t_l)$ of time-stamped edges a \emph{causal path} $p=\vv{v_0 v_1 v_2 \ldots v_l}$ of length $l$ from vertex $v_0$ to $v_l$ iff (i) $(v_i, v_{i+1}; t_i) \in E^{(t)}$, and (ii) $0 < t_{i+1} - t_i \leq \delta$ holds for $i \in \{0, 1, \ldots, l-1\}$.
We thus define the length of causal paths as the number of edges that they traverse, which implies that time-stamped edges are trivial causal paths of length one.
In this definition, the condition $0 < t_{i+1} - t_i$ ensures that the sequence of time-stamped edges respects the ``arrow of time'', while the condition $t_{i+1} - t_i \leq \delta$ ensures that to form a causal path two time-stamped edges must occur within time $\delta$ of each other.

We note that the existence of a causal path $\vv{v_0 \ldots v_l}$ is a necessary condition for a vertex $v_0$ in a dynamic graph to \emph{causally influence} another vertex $v_l$.
We further observe that each causal path in $G^{(t)}$ necessarily implies that the same path exists in the time-aggregated graph $G$.
Conversely, the existence of a path in graph $G$ corresponding to the dynamic graph $G^{(t)}$ does \emph{not} imply that the corresponding causal path exists in $G^{(t)}$.
The example in \Cref{fig:toyexample} (a) illustrates how the chronological order of edges can break transitivity in a dynamic graph.
Here, the timing and ordering of time-stamped edges implies that only two of the four theoretically possible \emph{causal paths} of length two exist. 
Hence, despite the presence of corresponding paths in the static topology, vertices $a$ and $b$ cannot indirectly influence $d$ and $e$ via causal paths $\vv{acd}$ and $\vv{bce}$ respectively.

To address the issue that time-aggregated graph representations discard information on causal paths, we utilize higher-order graph models that capture how the chronological ordering of edges influences causal paths~\cite{lambiotte2019}.
For a given dynamic graph $G^{(t)}$ and order $ k\geq 1$ we define a \textit{higher-order graph} $G^{(k)}$ as tuple $G^{(k)} = (V^{(k)}, E^{(k)})$ of higher-order vertices $V^{(k)} \subseteq V^k$ and higher-order edges $E^{(k)} \subseteq V^{(k)} \times V^{(k)}$.

Each higher-order vertex $v=:\vv{v_0 v_1 \ldots v_k} \in V^{(k)}$ is an ordered tuple of $k$ vertices $v_i \in V$ in the dynamic graph $G^{(t)}$ that also satisfies the conditions of a causal path.
Higher-order edges are constructed using the iterative line graph construction of high-dimensional De Bruijn graphs~\cite{deBruijn1946}.
The construction of a De Bruijn graph of order $k$ restricts edges to connect higher-order vertices that overlap in $k-1$ vertices, i.e. we require:

\[
(\vv{v_0 v_1 \ldots v_{k-1} v_{k}}, \vv{w_0 w_1 \ldots w_{k-1} w_{k}}) \in E^{(k)} \Rightarrow v_i = w_{i-1}(i=1, \ldots, k)
\]

Utilising the modelling framework introduced in \cite{Scholtes2017} we use (weighted) higher-order edges of a $k$-th order graph $G^{(k)}$ to represent the frequency of causal paths of length $k$ in a dynamic graph, i.e we define weights $w: E^{(k)} \rightarrow \mathbb{N}$ as 

\begin{align*}
w(\vv{v_0 \ldots v_{k-1}}, \vv{v_1 \ldots v_{k}}) & :=  \{ | (t_0, \ldots, t_{k-1}): (e;t_i) \in E^{(t)} \\ &\text{from causal path } \vv{v_0 \ldots v_k}) \}|
\end{align*}

\Cref{fig:toyexample} (c) shows an example for a (trivial) higher-order graph model of order $k=2$ that represents the causal paths $\vv{bcd}$ and $\vv{ace}$ of length two in the dynamic graph in \Cref{fig:toyexample} (a).
Higher-order graphs generalise time-aggregated graph representations of dynamic graphs, where for $k = 1$ we have $V^{(1)} = V$ and $E^{(1)} = E$.
Hence, a weighted time-aggregated graph is a first-order model of a dynamic graph that counts edges, i.e. causal paths of length one.
For $k>1$, we obtain higher-order models that capture both the topology and the chronological ordering of time-stamped edges in a dynamic graph, where the second-order model is the simplest model that is sensitive to the timing and ordering of edges.

{\bf Related Work}
\label{sec:background:graphDrawing}
Having illustrated the effects that are due to the arrow of time, we review works on dynamic graph drawing.
We only present methods relevant to our work, referring the reader to \cite{Beck2017_Taxonomy} for a detailed review.

A natural approach to visualise time series data on graphs are animated visualisations that show the temporal evolution of vertices and/or edges.
To generate such animations, we need to create a sequence of graphs, where each graph is a \emph{static snapshot} of the vertices and edges at one point in time.
An independent visualisation of such snapshots by means of standard graph layout algorithms is likely to result in animations that make it difficult to associate structures in subsequent frames, a problem often framed as maintaining the user's ``mental map''~\cite{purchase2006important,ArchambaultP12}.
A large number of works focuses on optimising graph layouts across multiple snapshots~\cite{diehl2001,erten2003simultaneous,gorg2004,kumar2006,loubier2008}, or in generating smooth transitions~\cite{nesbitt2002,friedrich2000} that minimise the cognitive effort required to trace time-varying vertices, edges, or clusters through subsequent snapshots.

Despite these improvements, identifying patterns in animations remains challenging.
Also, their use is limited since animations cannot be embedded in scholarly articles, books, or posters.
Addressing these issues, a second line of research focuses on methods to visualise dynamic graphs in terms of \emph{timeline representations}, which map the time dimension of dynamic graphs to a (static) spatial dimension.
Examples includes directed acyclic time-unfolded graph representations of dynamic graphs~\cite{Kempe2000_TemporalNetworks,Pfitzner2013_prl}, \emph{time arc} or \emph{time radar trees}~\cite{greilich2009,burch2012}, sequences of layered adjacencies~\cite{Vehlow2013}, stacked 3D representations where consecutive time slices are arranged along a third dimension~\cite{erten2003simultaneous}, and circular representations~\cite{van2014}. 
However, timelines are limited to a small number of time stamps, which hinders their application to data with high temporal resolution (e.g. seconds or even milliseconds).
The application of static graph drawing algorithms to such data requires a coarse-graining of time into \emph{time slices}, such that each time slice gives rise to a graph snapshot that can be visualised using, e.g., force-directed layout algorithms.
As pointed out in \cite{SimonettoAK17}, this coarse-graining of time leads to a loss of information on temporal patterns.
Some recent works have thus explored dynamic graph visualisations that are not based on time slices, e.g. using a continuous space-time cube~\cite{SimonettoAK17} or using visualisations that highlight higher-order dependencies at the level of individual nodes~\cite{tao2017}.
To the best of our knowledge none of the existing methods has explicitly addressed static representations of dynamic graphs that retain information on which nodes can influence each other via causal paths, which is the motivation for our work.

\begin{algorithm}[!ht]
\algtext*{EndFor}
\algtext*{EndIf}
\algtext*{EndWhile}
\algtext*{EndProcedure}
\caption{HOTVis: Higher-order time-aware layout with max. order $K$}

\label{alg:layout}
\begin{algorithmic}[1]
\Procedure{HOTVis}{$G^{(t)}, K, N, \delta, \alpha_2, \ldots, \alpha_K$}
	\State $A, \text{Pos} = dict(),\text{Temp}=t_0$
	\For {$k \in \text{range}(1,K)$}  \LineComment{superimpose attractive forces}
		\State $G^{(k)} = \text{HigherOrderGraph}(G^{(t)}, \delta, k)$
		\For {$(\vv{v_0 \ldots v_{k-1}}, \vv{v_1 \ldots v_k}) \in E^{(k)}$}		
			\If { $(v_0,v_k) \in A$}
				\State ${A[v_0,v_k] = A[v_0,v_k] + \alpha_k \cdot w(\vv{v_0 \ldots v_{k-1}}, \vv{v_1 \ldots v_k})}$
			\Else
				\State ${A[v_0, v_k] = \alpha_k \cdot w(\vv{v_0 \ldots v_{k-1}}, \vv{v_1 \ldots v_k})}$
			\EndIf
		\EndFor
	\EndFor
    \For {$i \in \text{range}(N)$} \LineComment{apply many-body simulation \cite{fruchterman1991}}
    	\For {$v \in V$}
        	\State $\Theta = 0$
            \For {$w \in V, w \neq v$}
                	\State $\Delta = \text{Pos}[w] - \text{P}[v]$
                    \State $\Theta = \Theta - \Delta/|\Delta| \cdot k^2/|\Delta|$
            \EndFor        
			\For {$(v,w) \in A$}
		    	\State $\Delta = \text{Pos}[w] - \text{Pos}[v]$
		    	\State $\Theta = \Theta + \Delta/|\Delta| \cdot A[v,w] \cdot |\Delta|^2/k$
		    \EndFor		
        \State $P[v] = P[v] + \Theta/|\Theta| \cdot \text{min}(|\Theta|, \text{Temp})$
       	\EndFor
    	\State $\text{Temp} = \text{cool}(\text{Temp})$
    \EndFor
    \Return \text{Pos}
\EndProcedure
\end{algorithmic}
\end{algorithm}

\section{Higher-Order Time-Aware Network Visualisation}
\label{sec:method}
To address the research gap outlined in \Cref{sec:background}, we propose an algorithm to generate higher-order time-aware visualisations (\texttt{HOTVis}). 
It captures the influence of the temporal dimension of a graph on its causal topology, i.e. which vertices can influence each other via causal paths, generalising the force-directed layout algorithm introduced in~\cite{fruchterman1991} to high-dimensional graphs.

Force-directed layouts optimally position vertices in a Euclidean space by means of a many-body simulation.
Attractive forces along edges move connected nodes close to each other while a repulsive force between all nodes separates them.
Simulating these forces until an equilibrium state is reached leads to graph layouts that highlight topological structures and symmetries~\cite{Battista1994}.
\texttt{HOTVis} generalises the attractive forces of force-directed layouts so that they capture the topology of causal paths in time-stamped data.
In particular, our algorithm \emph{superimposes} attractive forces that act between the endpoints of edges in multiple higher-order graphs up to a configurable maximum order $K$.
\Cref{fig:toyexample} (d) illustrates this idea based on the edges in the second-order model shown in \Cref{fig:toyexample} (c).
The additional attractive forces between vertex pairs $a,e$ and $b,d$ (coloured lines in \Cref{fig:toyexample} (d)) change the positioning of vertices such that those vertices that can causally influence each other are positioned in proximity.

The pseudocode of \texttt{HOTVis} is shown in \Cref{alg:layout}.
It takes a dynamic graph $G^{(t)}$, a maximum time difference $\delta$ used to define causal paths, a maximum order $K$, a number of iterations $N$, and parameters $\alpha_k$ controlling the influence of paths of length $k$ on the layout.
The algorithm works in two phases.
The first phase generates higher-order graphs $G^{(k)}$ up to order $K$ (lines 4--6). 
For each edge $(\vv{v_0 \ldots v_{k-1}}, \vv{v_1 \ldots v_k})$ in $G^{(k)}$, an attractive force is added between vertices $v_0$ and $v_k$ that can influence each other via causal path $\vv{v_0 v_1\ldots v_k}$ (lines 7--10).
Its strength depends on (i) the frequency of causal paths, and (ii) a parameter $\alpha_k$ that controls the influence of causal paths of length $k$ on the generated time-aware layout.
For $\alpha_2=\ldots=\alpha_K=0$ we obtain a standard force-directed (first-order) layout in which time is ignored.
For $\alpha_k>0$ and $k>1$ vertex positions are additionally influenced by the ordering of time-stamped edges.
The second phase of \texttt{HOTVis} (lines 11--22), uses the many-body simulation proposed in~\cite{fruchterman1991} to simulate repulsive and superimposed attractive forces between nodes.
The algorithm returns a dictionary of vertex positions that produces a static, time-aware visualisation.
The computational complexity of the algorithm is given by the sum of the computational complexities of two phases, the first consisting in the generation of $k-$th order graph models for $k = 2, \dots,K$, the second in the layouting of the nodes. 
The complexity of the first phase has a non-trivial dependency on the temporal distribution of time-stamped edges of the dynamical graph and is further discussed in the appendix.
The computational complexity of the second phase corresponds to that of the algorithm proposed in \cite{fruchterman1991}.

\begin{figure}[!ht]
     \centering
         \includegraphics[width=.325\textwidth]{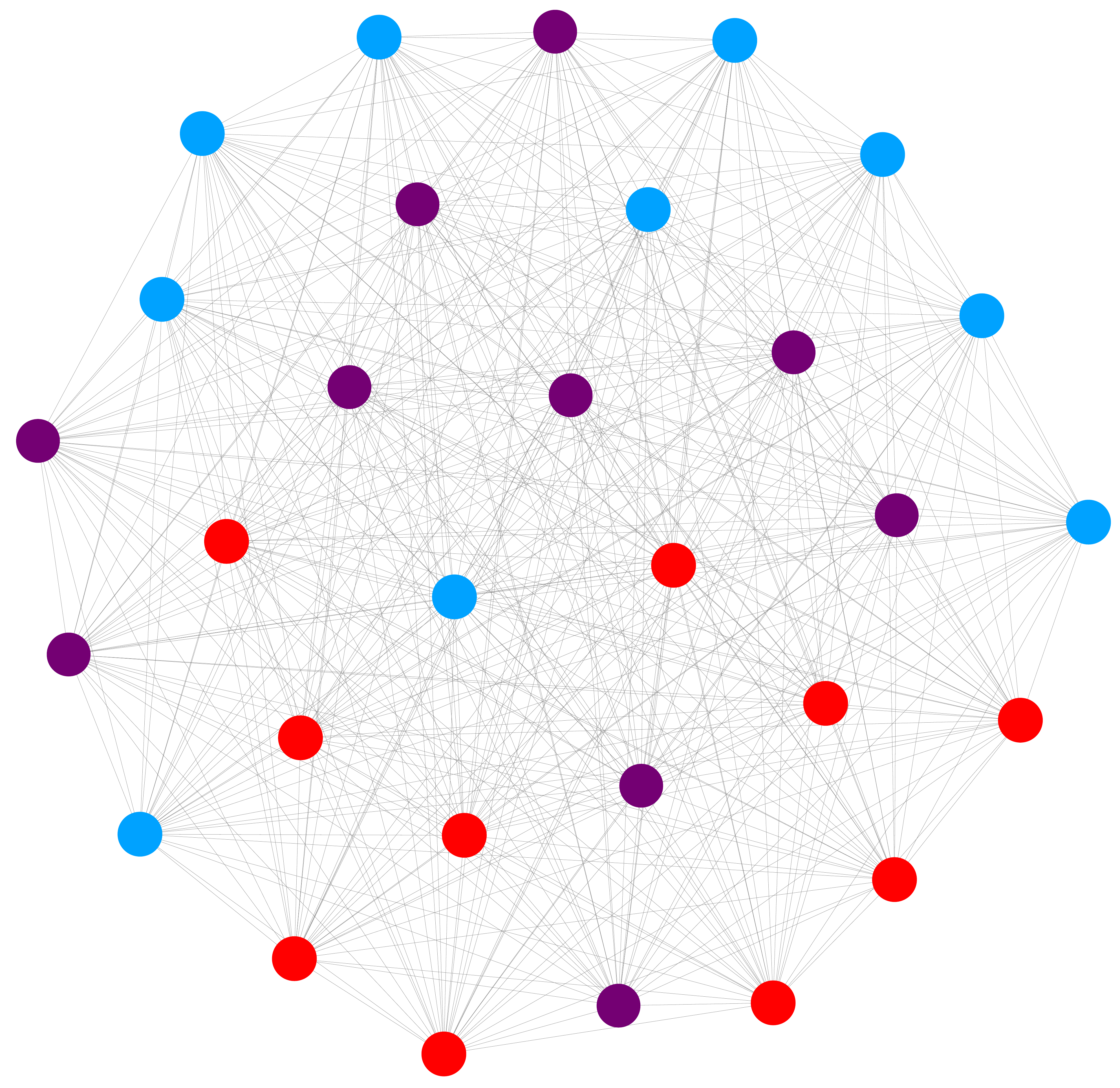}
         \includegraphics[width=.325\textwidth]{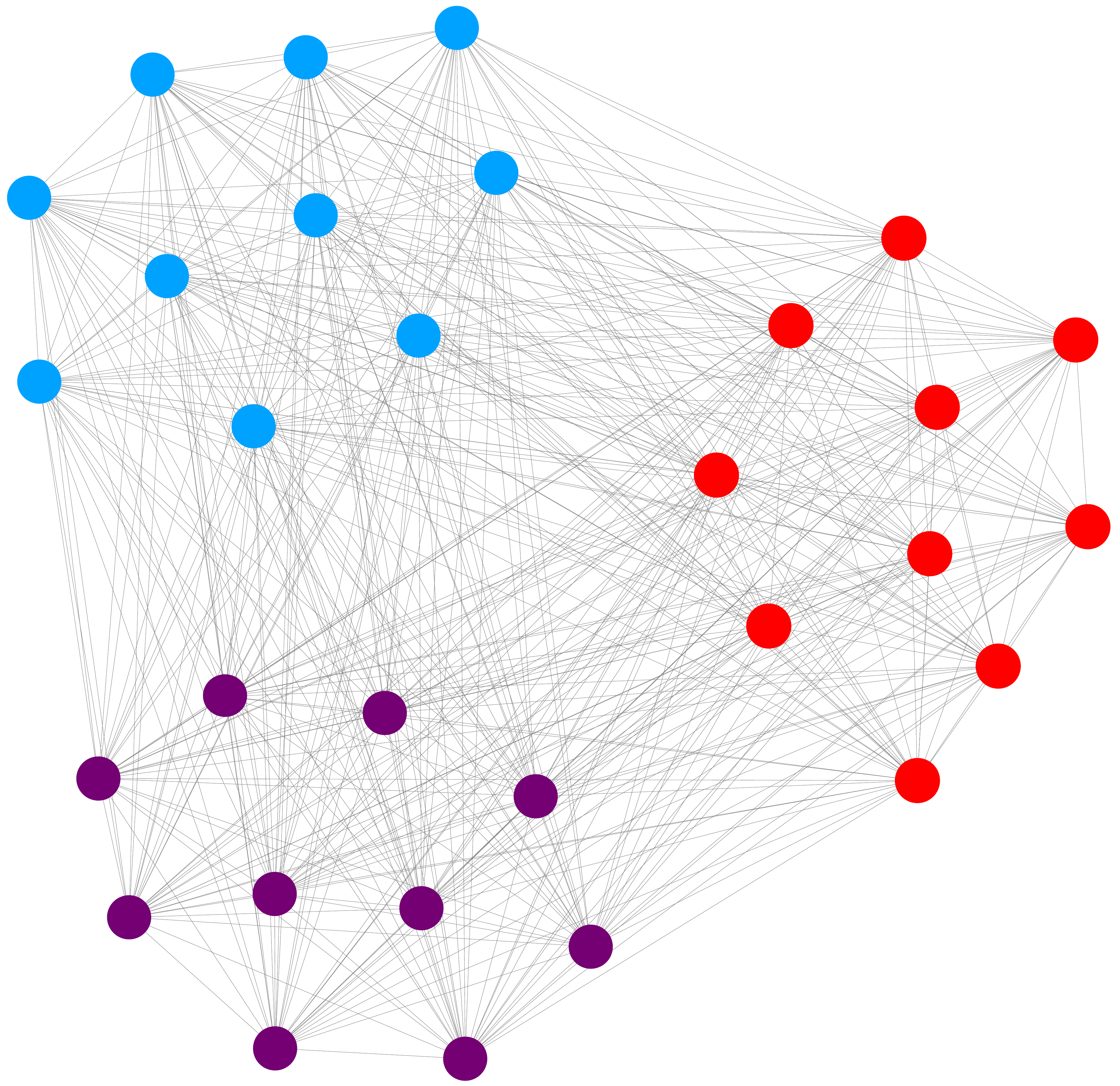}
         \includegraphics[width=.325\textwidth]{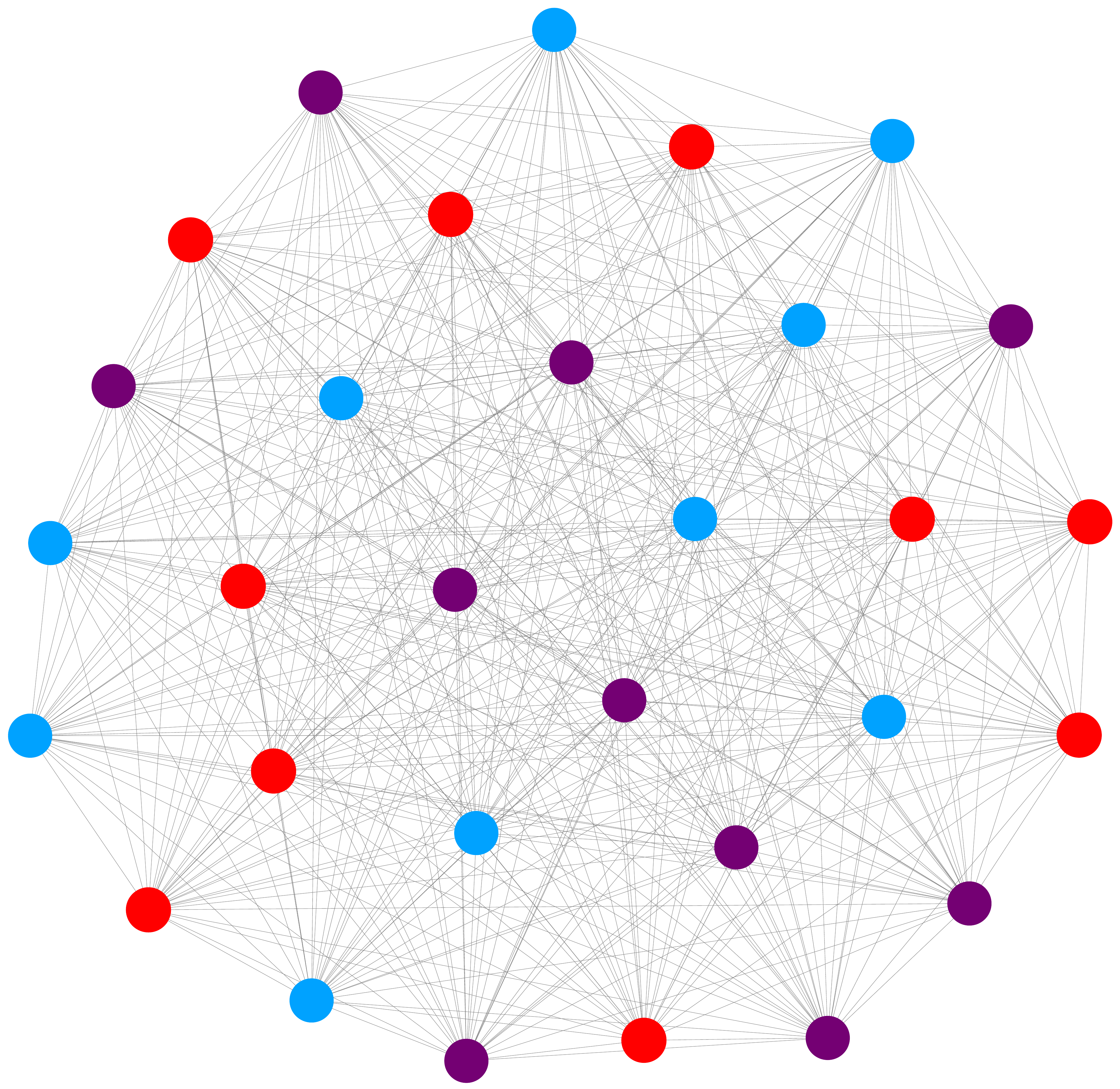}
        \caption{Application of HOTVis to synthetic temporal network with three temporal clusters (coloured nodes). A second-order time-aware layout (middle) highlights clusters not visible in a static visualisation (left). A time-aware visualisation of the data with randomised time stamps (right) confirms that clusters are due to the ordering of edges.}
        \label{fig:synth}
\end{figure}

{\bf Illustration in synthetic example} A demonstration of a time-aware visualisation in synthetic data on a temporal network with $K=2$ is shown in \Cref{fig:synth}.
The data was generated using a stochastic model creating time-stamped edge sequences with temporal correlations that lead to an over-representation of causal paths of length two that (indirectly) connect pairs of vertices in three clusters (coloured vertices in \Cref{fig:synth}).
Different from what one would expect based on the time-aggregated topology, the chronological ordering of time-stamped edges leads to an under-representation of causal paths between vertices with \emph{different colours}.
Hence, we obtain \emph{temporal clusters in the causal topology}, where vertices in the same cluster can indirectly influence each other via causal paths more than vertices in different clusters.
Details of the model are included in the appendix and in a Zenodo package (cf. reproducibility notes in appendix).
\emph{Temporal clusters} in the causal topology are visible only in a second-order time-aware visualisation that superimposes attractive forces calculated in both the first and the second-order graph (middle panel \Cref{fig:synth}).
These clusters are not visible in the static graph layout shown in \Cref{fig:synth} (left panel).
To demonstrate that clusters are solely due to the chronological ordering of time-stamped edges and the resulting causal paths, we shuffle the timestamps of edges and reapply our time-aware visualisation algorithm.
The resulting layout in \Cref{fig:synth} (right panel) shows that the shuffling of timestamps destroys the cluster structure, confirming that \texttt{HOTVis} visualises patterns due to the ordering of edges.

\section{Experimental Evaluation}
\label{sec:results}

Having illustrated \texttt{HOTVis} in a synthetic example, we now evaluate whether it produces better visual representations of empirical time-stamped network data.
We compare \texttt{HOTVis} (at different orders) against a baseline visualisation that is generated using the Fruchterman-Reingold algorithm~\cite{fruchterman1991}.
To quantitatively assess the ``quality'' of visualisations generated by \texttt{HOTVis}, we define measures that capture how well the causal topology is represented.
For the following definitions, let $G^{(t)}$ be a dynamic graph that gives rise to a multi-set $S$ of $N$ causal paths $S= \{ p_0, \ldots, p_N \}$.
We also assume that \Cref{alg:layout} assigns each vertex $v \in V$ to a position $\pi_v:=Pos[v] \in \mathbb{R}^2$.

\textbf{Edge crossing ($\xi$)} 
This standard measure counts the number of pairs of edges that cross each other in the visualisation.
It is widely used in the evaluation of graph drawing algorithm. 
It rests on the idea that a large number of edge crossings $\xi$ makes it difficult to identify which vertices are connected by edges, i.e. ``high-quality'' drawings minimise $\xi$.
We efficiently calculate edge crossing based on the orientation predicate~\cite{Shewchuk1997_ComputationalGeometry}.

\textbf{Causal path dispersion ($\sigma$)}
The \emph{causal path dispersion} $\sigma$ captures whether the sets of vertices traversed by causal paths are less spatially dispersed than expected based on the spatial distribution of vertices.
It intuitively captures whether vertices that can influence each other directly and indirectly are positioned in close proximity.
For this, we consider a multi-set $S$ of causal paths $p$ with cardinality $N := |S|$ that traverse a graph with vertices $V$. 
We define $\sigma$ as

\[ \sigma =  \frac{\sum_{p \in S} \sum_{u_i \in p}||Pos[u_i] - B(p)||\cdot |V|}{N \cdot \sum_{v_i \in V}|| Pos[v_i] -  B(V)||} \in \left[0, 1\right], \]

where $B: 2^V \rightarrow \mathbb{R}^2$ is a function that returns the barycentre of vertex positions $P[v]$ for vertex multi-set $V$.
For $\sigma \approx 1$ the spatial distribution of nodes traversed by causal paths is the same as for nodes traversed by random paths in the network topology.
Values of $\sigma < 1$ highlight that vertices connected via causal paths occupy a smaller area than expected at random.

 \textbf{Closeness eccentricity ($\Delta$)}
In force-directed layouts the distance of a vertex from the barycentre of the visualisation is correlated with the vertex' \emph{closeness centrality}, defined as the inverse of the sum of shortest path lengths between the vertex and all other vertices~\cite{Brandes2003}.
It is thus interesting to study whether vertex positions in our time-aware visualisation are correlated with the \emph{temporal closeness centrality} of a vertex $v$. For a set $S$ of causal paths $p$, we define this as
 
\[ CC(v) := \sum_{w \neq v \in V}\frac{\sum_{p\in S} \delta_w(p)\delta_v(p)}{\sum_{p\in S, w \in p} \text{dist}(v,w;p)}. \]

Here, $\text{dist}(v,w;p)$ denotes the (topological) distance between vertices $v$ and $w$ via causal path $p$ and $\delta_v(p)$ is one if path $p$ traverses vertex $v$ and zero otherwise.
With $T_{\gamma} = \{u_1, u_2, \ldots, u_n\}$ being the set of $n$ nodes whose temporal closeness centrality is in the $\gamma$ upper percentile, we define closeness eccentricity $\Delta(\gamma)$ as:

\[
\Delta(\gamma) := \frac{\sum_{u_i \in T_{\gamma}} ||Pos[u_i]-B(V)|| \cdot |V|}{|T_{\gamma}| \cdot \sum_{v \in V} ||Pos[v]-B(V)||} \in \left[0, 1\right]
\]

$\Delta(\gamma)$ captures whether the $n$ vertices with highest temporal closeness centrality are closer to ($\Delta<1$) or farther away ($\Delta>1$) from  the barycentre of the visualisation than we would expect at random.

\subsection{\textbf{Experimental results}} 
We now report the results of our experimental evaluation of \texttt{HOTVis} in (i) the synthetically generated data with temporal clusters introduced above, and (ii) five time-stamped data sets on real complex networks.
The five data sets fall into two classes, highlighting different types of data in which our algorithm can be used. 
The first class captures paths or trajectories in a networked system. 
Here we utilise two data sets (i) \texttt{flights}, which captures $280k$ passenger itineraries between $175$ US airports recorded in $2001$, and (ii) \texttt{tube}, which contains $4.2$ million passenger trips in the London metro in $2014$. 
Details on those data are available on Zenodo \cite{zenodo}.
The second class consists of time-stamped data on social interactions, in which we can calculate \emph{causal paths} as defined in \cref{sec:background}.
Here we used three data sets from the Sociopatterns collaboration, namely (i) \texttt{hospital}, which captures 32,424 time-stamped proximity events between 75 patients, medical and administrative staff in a hospital recorded over a period of five days~\cite{Vanhems2013_TransmissionRoutes}, (ii) \texttt{workplace}, which consists of 9,827 face-to-face interactions between 92 company employees recorded in an office building over a period of ten days~\cite{Genois2015_FaceToFace} and (iii) \texttt{school} which contains 77,602 proximity events between 242 individuals (232 children and 10 teachers) \cite{stehle2011high}.

We now tests whether we can ``learn'' consistent patterns in the causal topology of the time-stamped network data from the time-aware visualisation generated by \texttt{HOTVis}.
To assess the consistency of the patterns identified by our algorithm we use a cross-validation approach: we generate time-aware visualisations with different maximum orders $K$ in a training sample, and then assess their quality in a validation set that we withheld from our algorithm.
Thanks to the evaluation on unseen data, we can use our measures to compare the generalisability of the patterns displayed at each maximum order $K$.
A benefit of our method is that the optimal maximum order needed to visualise the topology of causal paths can be determined using the statistical model selection techniques described in~\cite{Scholtes2017,petrovic2020learning}.
It provides a principled method to balance the complexity and explanatory power of the higher-order model used for our visualisation, learning a model that avoids both over- and underfitting.
In the following, we evaluate our algorithm for all maximum orders $K=1, \ldots, K_{opt}+2$, where $K_{opt}$ is the \emph{optimal} order returned by the method described in~\cite{Scholtes2017}.
All results were obtained by applying the cross-validation approach described above in $100$ layout calculations for each data set and for each maximum order $K$.
To focus on the effect of the \emph{topology} of causal paths and rule out distortions due to skewed distributions of weights $w$ (cf. \cite{jacomy2014}), we set the weight of all paths to a constant value of one.
Moreover, to ensure that all orders have an equal influence on the overall layout, we set the parameter $\alpha_k$ in a higher-order model with order $k$ to $\alpha_k := m_k^{-1}$, where $m_k$ is the number of unique paths of length $k$.
This ensures that the forces in each $k$-th order model are scaled according to the density of edges.
We note that the choice of those parameters is motivated by simplicity and ease of reproducibility. 
In particular, it does not require sophisticated parameter tuning, which could be used to optimise the visualisation for a specific data set.

\begin{figure*}[!ht]
    \centering
	\includegraphics[width=\textwidth]{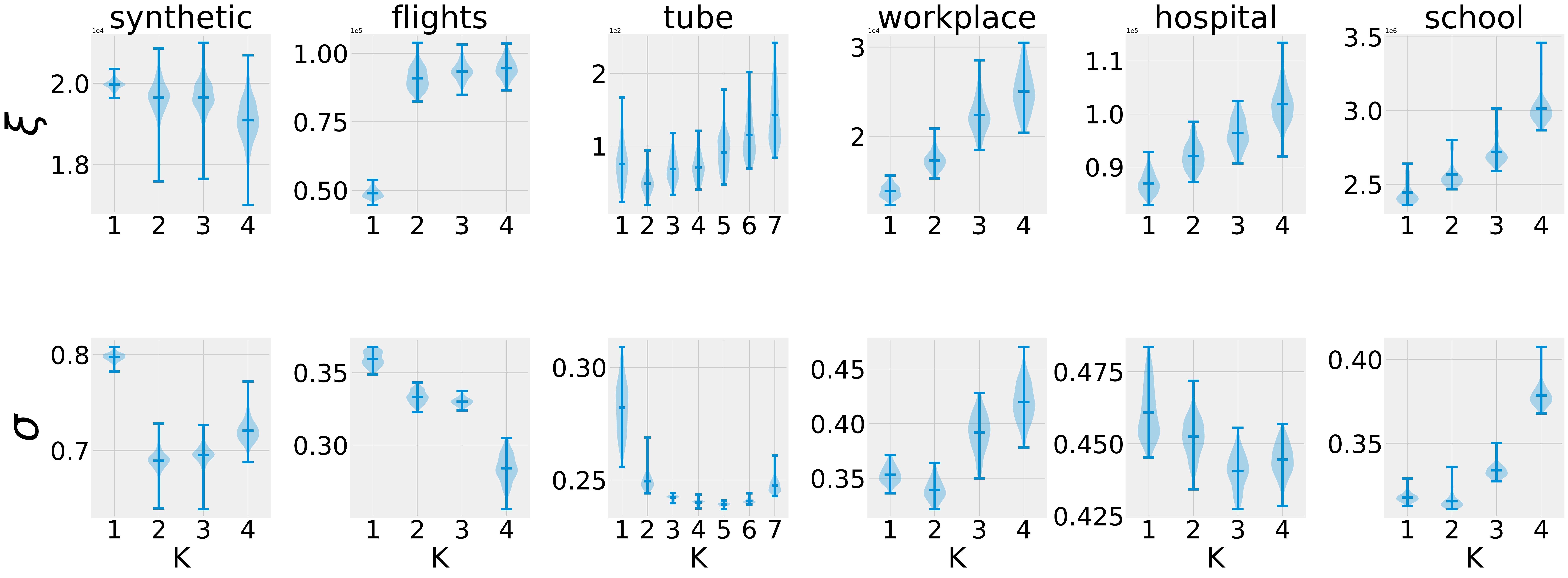}
    \caption{Edge crossing $\xi$ (top) and causal path dispersion $\sigma$ (bottom) for a synthetic dynamic graph with three clusters (column \texttt{synthetic}), empirical data on flight itineraries (column \texttt{flights}), metro trips (\texttt{tube}), and time-stamped interactions between workers in an office environment (\texttt{workplace}), patients and hospital staff (\texttt{hospital}), and children and teachers in a primary school (\texttt{school}). All values are averages of $100$ cross-validation experiments, where a time-aware layout with maximum order $K$ (x-axis) was computed for a $70 \%$ random training set of causal paths, calculating quality measures (y-axis) in the layout for a test set of remaining $30 \%$ of causal paths.}\label{fig:res}
\end{figure*}

The results of our evaluation are shown in \Cref{fig:res}.
For the edge crossing $\xi$, in the synthetic data we find no significant change with increasing $K$, while the empirical data sets show significant increases as the maximum order grows.
A general growth of edge crossings with increasing $K$ is expected since, apart from the topology of \emph{edges}, the time-aware visualisation considers the topology of causal paths.
The causal path dispersion $\sigma$ decreases considerably for orders $K>1$ in all data sets, highlighting that our algorithm positions those nodes close to each other that strongly influence each other via causal paths.
For a suitably chosen order $K$, we further observe that relatively large decreases of causal path dispersion $\sigma$ (e.g. a decrease of $15 \%$ for order $K=5$ in \texttt{tube}) are associated with relatively mild or insignificant increases of edge crossing $\xi$ (e.g. no significant change for order $K$ in \texttt{tube}).
For those orders $K$, our method provides a good trade-off between a visualisation that best represents the topology of causal paths and a visualisation that best represents the static topology.
On the one hand, this supports our hypothesis that \texttt{HOTVis} better represents the causal topology of temporal networks compared to time-aggregated (first-order) visualisations.
On the other hand, this raises the issue of finding the ``optimal'' order $K$ of a higher-order graph model, which can be addressed using the statistical model selection technique presented in~\cite{Scholtes2017}. 
In agreement with the results of our cross-validation, this technique yields an optimal order $K_{opt}=5$ for \texttt{tube} and $K_{opt}=2$ for the other data sets.
This indicates that we can use statistical model selection to learn the optimal maximum order parameter $K$ to be used in \texttt{HOTVis}.

\begin{figure}[!ht]
	\centering{}
			\includegraphics[width=.49\textwidth]{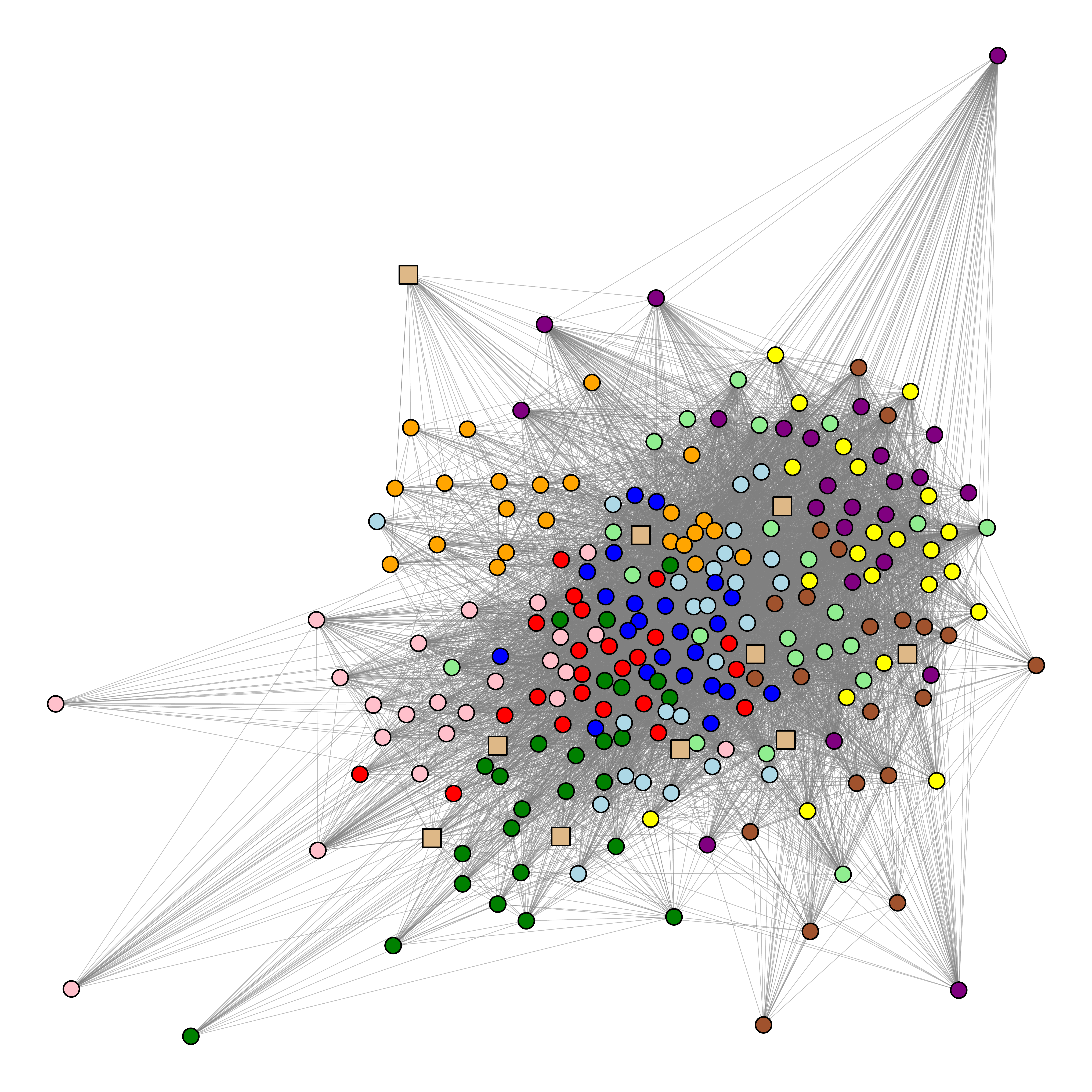}
			\includegraphics[width=.49\textwidth]{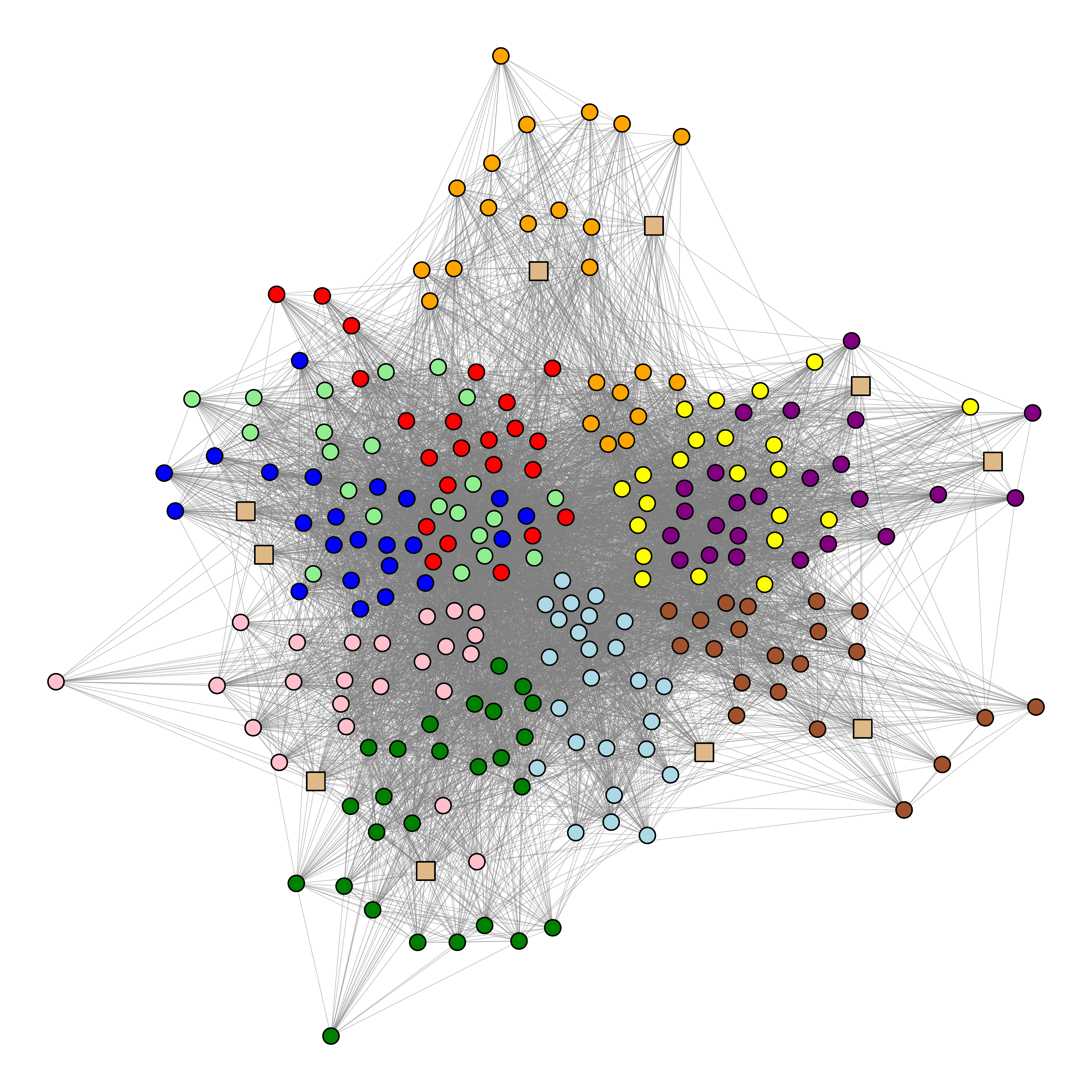}
		\caption{Comparison between a time-neglecting (left) and time-aware (right) layout for the \texttt{school} data set. Node colours represent the class each student belongs to, while square markers identify teachers. The time-aware visualisation generated by \texttt{HOTVis} (right) positions nodes that influence each other through causal paths close to each other. 
		\texttt{HOTVis} highlights ground truth groups of students in school classes. We used the default parameter $\alpha_k=m_k^{-1}$ as described in \Cref{sec:results}.}\label{fig:res:example_school}
	\end{figure}

In \Cref{fig:res:example_school} we illustrate \texttt{HOTVis} in the \texttt{school} data set for $K = 1$ (left) and the optimal order $K = 2$ (right).
Node colours indicate the membership of students in different classes.
Importantly, this group structure in the data is not expressed in the topology of links (see \Cref{fig:res:example_school} left).
Consequently, a time-neglecting first-order layout places nodes in a single group, which leads to a cluttered visualisation that makes it difficult to visually detect the ground truth group structure.
A second-order layout generated by \texttt{HOTVis}  (see \Cref{fig:res:example_school} right) better highlights group structures that are expressed in the topology of causal paths, thus leading to \emph{temporal cluster} patterns that cannot be seen in the static topology.
This example demonstrates that the mechanism by which \texttt{HOTVis} visualises temporal clusters---as illustrated in the synthetic example in \Cref{fig:synth}--- can successfully visualize group structures in empirical social networks.

\begin{figure*}[!ht]
	\centering
		\centering
	    \includegraphics[width=\textwidth]{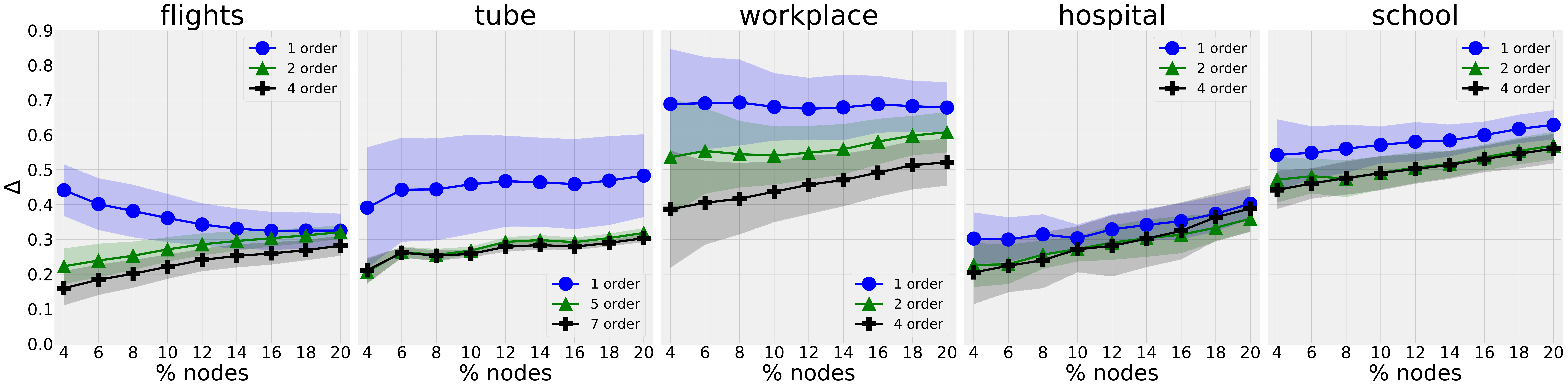}
	    \caption{Closeness eccentricity $\Delta$ (y-axis) for a varying top percentage $n$ of vertices with highest temporal closeness (x-axis). 
	    Results are presented for $K$=$1$ (blue dots) $K$=$K_{opt}$ (green triangles), $K$=$K_{opt+2}$ (black pluses). Hulls indicate the 2$\sigma$ interval.}
		\label{fig:res:centrality}
	\end{figure*}

\begin{figure*}[!ht]
	\centering
	\includegraphics[width=\textwidth]{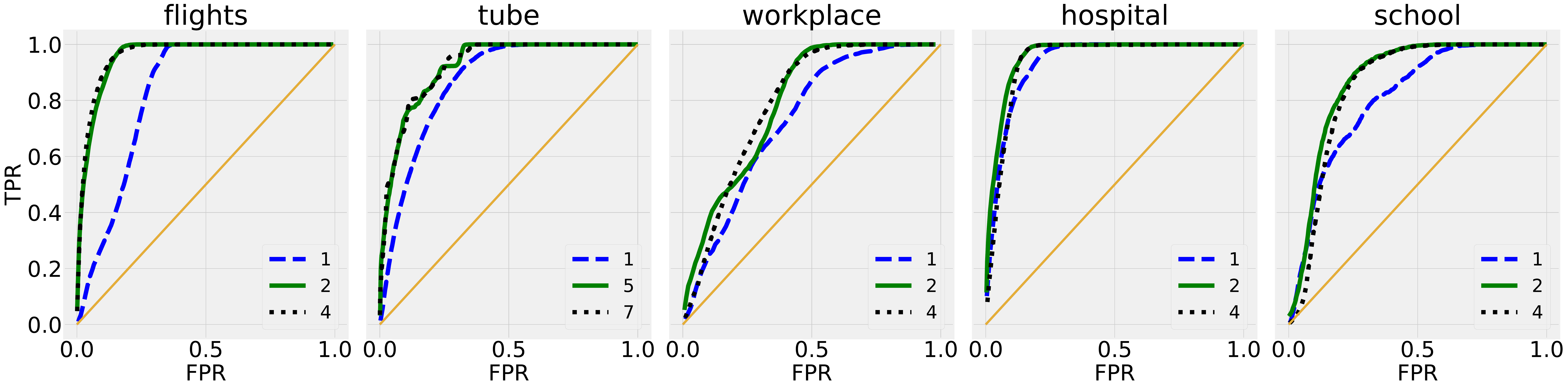}
	\caption{ ROC curves illustrate the ability to predict nodes with top $10\%$ closeness centrality for $K$ = $1$ (blue dashed), $K$=$K_{opt}$ (green solid), $K$=$K_{opt+2}$ (black dotted).}
	\label{fig:res:roc_center}
\end{figure*}

\begin{figure}[!ht]
	\centering{}
			\includegraphics[width=.49\textwidth]{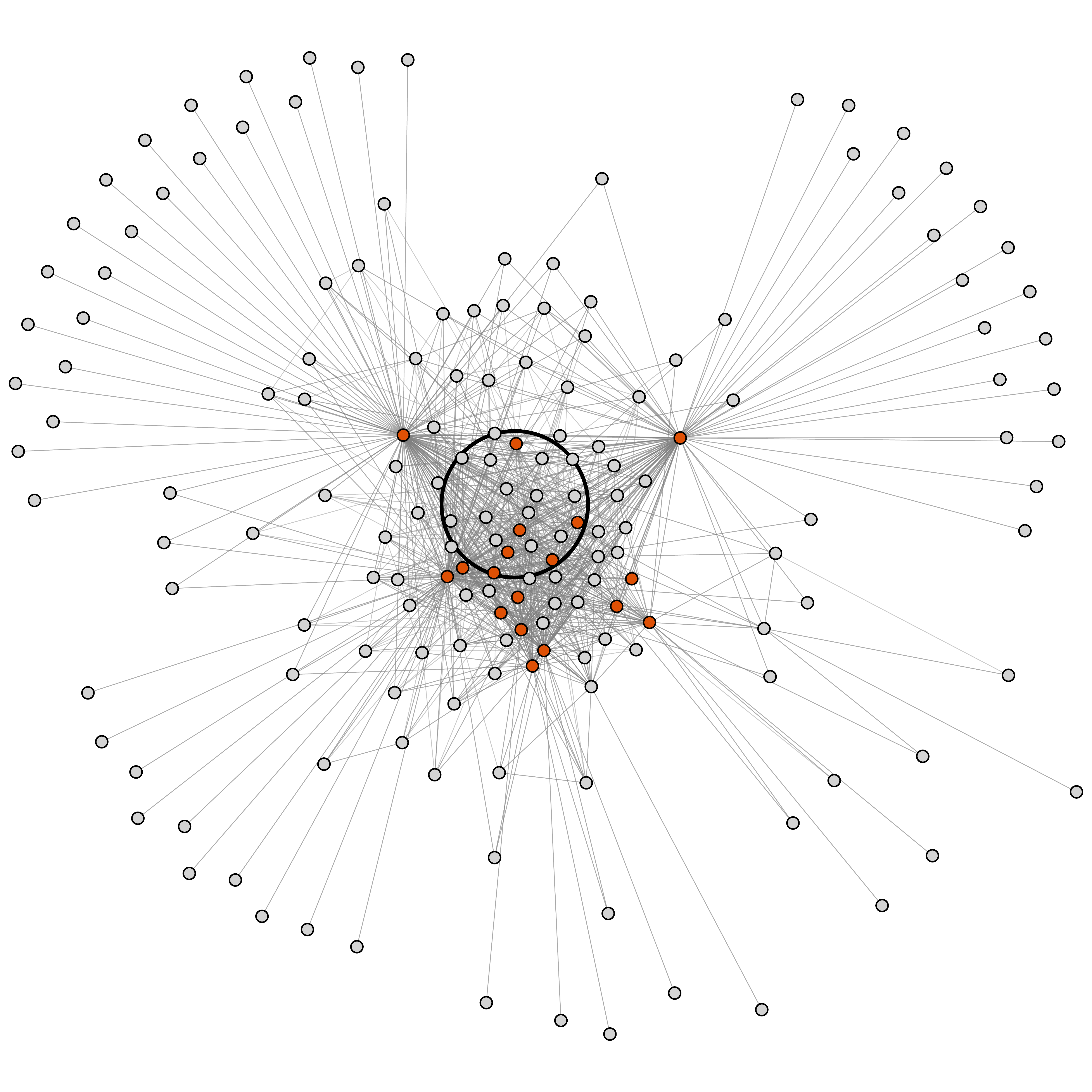}
			\includegraphics[width=.49\textwidth]{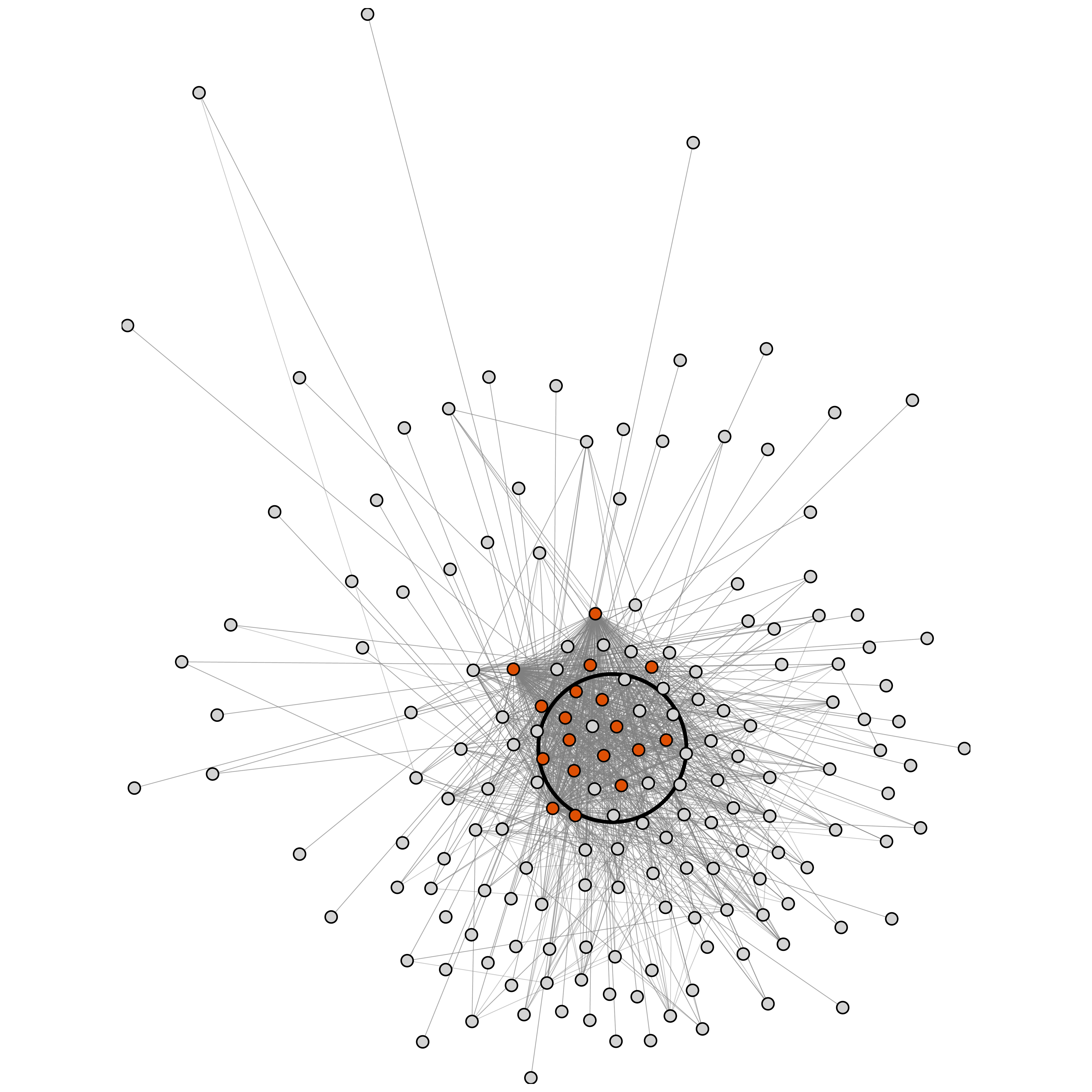}
		\caption{Comparison between time-neglecting (left) and time-aware (right) layout for \texttt{flights}. Vertices whose temporal closeness centrality is among top 10 \% of values are highlighted in red. Black circles delineate the area containing $10 \%$ of vertices closest to the barycentre. The time-aware visualisation generated by \texttt{HOTVis} (right) places vertices with high temporal closeness centrality close to the barycentre, enabling us to identify temporally important vertices. We used $\alpha_k=m_k^{-1}$ as described in \cref{sec:results}.}\label{fig:res:example}
	\end{figure}

{\bf Temporal closeness.} 
We finally show results for closeness eccentricity $\Delta$.
We specifically test whether, similar to static force-directed layout algorithms, \texttt{HOTVis} places nodes with high temporal closeness in the centre of the visualisation.
The results for the four empirical data sets are shown in \cref{fig:res:centrality} (again for $100$ cross-validation experiments).
In all data sets, higher values of $K$ correspond to lower values of $\Delta$.
For \texttt{tube} and \texttt{flights}, values significantly differ with a $2\sigma$ confidence interval.
This indicates that \texttt{HOTVis} represents the temporal closeness of vertices better than a first-order layout.
To quantify our ability to visually identify vertices with high temporal centrality, we additionally ran a prediction experiment:
We use the proximity of vertices to the barycentre of the visualisation to identify vertices whose temporal closeness is in the top $10 \%$ percentile.
This yields a binary classification problem, where we use vertex positions calculated by \texttt{HOTVis} in a training set to predict vertices with highest temporal closeness centrality in a validation set.
We predict a vertex to be among the top $10 \%$ vertices with highest temporal closeness centrality, if it is among the top $10 \%$ vertices with smallest distance to the barycentre. 
To compare the performance of different orders $K$ we use ROC curves ($100$ cross-validation experiments).
The results in \Cref{fig:res:roc_center} show that for $K>1$ \emph{HOTVis} outperforms a static (first-order) visualisation in all data sets.
AUC scores for $K = 1$, $K = K_{opt}$ and $K =K_{opt}+2$ are: $0.82$, $0.95$, $0.95$ in \texttt{flights}; $0.87$, $0.92$, $0.92$ in \texttt{tube}; $0.71$, $0.77$, $0.77$ in \texttt{workplace};  $0.93$, $0.96$, $0.94$ in \texttt{hospital}; $0.80$, $0.87$, $0.85$ in \texttt{school}.
Supporting our hypothesis, we find that visualisations with order $K>K_{opt}$, where $K_{opt}$ is determined by the model selection technique from~\cite{Scholtes2017}, only yield negligible increases (or even decreases) in prediction performance.
We illustrate this in the \texttt{flights} dataset for $K = 1$ and $K=K_{opt} = 2$.
The two layouts in \Cref{fig:res:example} strongly differ in the positioning of the top $10 \%$ vertices with highest temporal closeness centrality (in red).
Different from a time-neglecting layout, in \texttt{HOTVis} the majority of vertices classified as most central due to their proximity to the barycentre are within the top $10 \%$ values of temporal closeness centrality.


\section{Conclusion and Outlook}
\label{sec:conclusion}

Despite advances in dynamic graph drawing, the visualisation of high-resolution time-stamped network data is still a challenge.
Existing methods suffer from a limited ability to highlight patterns in the \emph{causal topology of dynamic graphs}, which is determined by the interplay between its topology (i.e. \emph{which} edges exist) and the temporal dynamics of edges (i.e. \emph{when} time-stamped edges occur).
We address this issue through \texttt{HOTVis}, a dynamic graph drawing algorithm that uses higher-order graph to produce static, time-aware visualisations.
Experiments in synthetic and empirical data support our hypothesis that the resulting visualisations better highlight temporal clusters due to the chronological ordering of edges.
We further show that \texttt{HOTVis} places highly influential vertices (i.e. vertices with high temporal closeness) close to the centre of the visualisation, which better represents their role in the system and supports visual data mining.

Our algorithm introduces an additional parameter ---the maximum order $K$ to be used for the visualisation--- that needs to be adjusted to the temporal correlations in the data.
We show that recent advances in statistical modelling and machine learning enable us to automatically learn the optimal choice $K_{opt}$ for this parameter, thus turning it into a practicable method to visualise patterns in temporal data. 
Our work highlights a largely unexplored potential for new visual data mining techniques that combine graph drawing, higher-order network models~\cite{lambiotte2019}, and machine learning that we seek to explore in the future.

\section*{Acknowledgements}
Vincenzo Perri and Ingo Scholtes acknowledge support by the Swiss National Science Foundation, grant 176938. Ingo Scholtes acknowledges support by the project bergisch.smart.mobility, funded by the German state of North Rhine-Westphalia.

\bibliographystyle{splncs04}
\bibliography{ref}

\clearpage

\renewcommand{\thesection}{\Alph{section}}

\begin{Large}\textbf{Supplementary Information}\end{Large}
\\
In this appendix, we provide supplementary information to (i) ensure the reproducibility of our experimental results, and (ii) further back up our claims regarding the benefits of higher-order time-aware layouts of dynamic graphs.

\section{Notes on Computational Complexity}
\label{sec:complexity}

We briefly comment on the computational complexity of \texttt{HOTVis}.
We first note that the complexity of the second phase of Algorithm 1 corresponds to the computational complexity of the well-known force-directed layout algorithm introduced in~\cite{fruchterman1991}.
The additional computational effort that is introduced in the first phase of our algorithm directly depends on the computational complexity of generating $k$-th order graph models for $k=2, \ldots, K$.
While we refer the reader to \cite{Petrovic2019_PaCo} for a detailed discussion (and proof) of the complexity of generating higher-order models of causal paths of length $k$, here we highlight that a term for the worst-case complexity can be given as 

\[ \mathcal{O} (N \cdot |V| \cdot K^2 \cdot [m\delta \lambda_{\text{max}}^{K-2} + \lambda_{\text{max}}]), \]

where $N$ is the number of time-stamped edges in the dynamic graph, $|V|$ is the number of vertices, $K$ is the maximum order of the higher-order graph model used in the time-aware visualisation, $m$ is the maximum number of time-stamped edges with the same time stamp, $\delta$ is the maximum time difference used in the definition of causal paths, and $\lambda_{\text{max}}$ is the largest eigenvalue of the binary adjacency matrix of the time-aggregated graph topology.

This result shows that the computational complexity of our visualisation algorithm strongly depends on the temporal distribution of time-stamped edges, which influences the number of causal paths of length $k$ and thus the size of a $k$-th order graph. 
Moreover, the scaling of computational complexity with the maximum order $K$ that is used in the visualisation depends on the sparsity of the time-aggregated topology expressed in the leading eigenvalue $\lambda_{\text{max}}$ of the adjacency matrix.

For empirical data, we find that higher-order graph models are generally highly sparse, which enables us to compute higher-order time-aware graph layouts up to a maximum order $K\approx 10$ in a few seconds even for data on dynamic graphs with millions of time-stamped edges.

\section{Model for Temporal Networks with Temporal Clusters}
\label{sec:model}

We provide additional details on the stochastic model used to generate synthetic dynamic graphs with temporal cluster structure and a random time-aggregated topology.
The model performs the following three steps:
\begin{enumerate}[topsep=0pt]
	\item Generate a static random $k$-regular graph with $n$ vertices, where each vertex is connected to a random set of $k$ neighbours. Randomly assign the $n$ vertices to three equally-sized, non-overlapping clusters, where $C(v)$ denotes the cluster of vertex $v$.
	\item Generate $N$ sequences of two randomly chosen time-stamped edges $(v_0,v_1;t)$ and $(v_1,v_2;t+1)$ that contribute to a causal path of length two in the resulting dynamic graph.
	\item For each vertex $v_1$ of such a causal path of length two randomly pick:
	\begin{itemize}
		\item two time-stamped edges $(u, v_1; t_1)$ and $(v_1, w, t_1+1)$ such that $C(u)=C(v_1) \neq C(w)$
		\item two time-stamped edges $(x, v_1; t_2)$ and $(v_1, z; t_2+1)$ with $C(v_1)=C(z) \neq C(x)$
	\end{itemize}
	\item Swap the time stamps of the four time-stamped edges to $(u, v_1; t_1)$ and $(v_1, z; t_1+1)$, $(x, v_1, t_2)$, and $(v_1, w, t_2+1)$.
\end{enumerate}

This simple procedure exclusively changes the temporal ordering of time-stamped edges, affecting neither the topology nor the frequency of time-stamped edges.
The model changes time stamps of edges (and thus causal paths) such that vertices are preferentially connected---via causal paths of length two---to other vertices in the same cluster.
This leads to a strong cluster structure in the causal topology of the dynamic graph, which (i) is neither present in the time-aggregated topology nor in the temporal activation patterns of edges, and (ii) can nevertheless be visualised by our time-aware graph visualisation algorithm (see figure 2 in the main text).

An interactive HTML animation of a dynamic graph generated by this model and an interactive demo of the resulting time-aware static visualisation are available online \footnote{see \url{http://www.pathpy.net/clustering.html}}.
The python code with as an interactive \texttt{jupyter} demonstration is available via the open research repository \texttt{zenodo.org} \cite{zenodo}.

\section{Software Implementation and Reproducibility}
\label{sec:repro}

The proposed algorithm \texttt{HOTVis}, as well as the quality measures have been implemented in the OpenSource \texttt{python} data analytics and visualisation package \texttt{pathpy}~\cite{pathpy}.
Interactive tutorials, animations, and illustrative examples that showcase our concept of higher-order time-aware layouts of dynamic graphs, and their implementation are available online at \url{www.pathpy.net}.
The empirical travel datasets are freely available at \cite{FLdata,LTdata}.
The empirical time-stamped social network data that were used in this work are freely accessible at \url{www.sociopatterns.org}.
Both the code and the data required to reproduce our results are available via the open research repository \texttt{zenodo.org} \cite{zenodo}.


\end{document}